\documentstyle[12pt,a4]{article}
\begin{document}
\title{
The rigidly rotating disk of dust and its black hole limit}
\author{
Reinhard Meinel\\Friedrich--Schiller--Universit\"at Jena\\
Theoretisch--Physikalisches Institut\\
Max--Wien--Platz 1, D--07743 Jena, Germany\\
E-mail: meinel@tpi.uni-jena.de
}
\date{}
\maketitle
\begin{abstract}
The exact global solution of the Einstein equations [Neugebauer \& Meinel, 
Phys.~Rev.~Lett.~75 (1995) 3046] describing a rigidly rotating,
self--gravitating disk is discussed. The underlying matter model is a perfect
fluid in the limit of vanishing pressure. The solution represents the
general--relativistic analogue of the classical Maclaurin disk. It was derived 
by applying solution techniques from soliton theory to the axisymmetric, 
stationary vacuum Einstein equations. In contrast to the Newtonian solution,
there exists an upper limit for the total mass of the disk -- if the
angular momentum is fixed. At this limit, a transition to a rotating black hole,
i.e., to the Kerr solution
occurs. Another limiting procedure leads to an interesting cosmological
solution. These results prove conjectures formulated by Bardeen and Wagoner  
more than twenty--five years ago.
\end{abstract}

\vspace{5cm}
\begin{center} {\it Proceedings of the \\Second Mexican School on Gravitation 
and Mathematical Physics.}

Edited by A.~Garcia, C.~L\"ammerzahl, A.~Macias, T.~Matos, 
D.~Nu\~nez.

Sciencia Verlag Konstanz, 1997
\end{center}
\thispagestyle{empty}

\newpage
\section{Introduction: \newline Rotating bodies in general relativity}
The problem of self--gravitating rotating bodies is a {\it global}\/ one: One
has to solve simultaneously the interior equations (with matter) 
as well as the exterior (vacuum) equations. The shape of the surface
cannot be prescibed arbitrarily. A famous solution within Newton's theory of
gravitation is the Maclaurin spheroid of a rigidly rotating perfect fluid with
constant mass density (cf.~\cite{li}, \cite{cha}). 
Within Einstein's theory of gravitation the problem is
even more complicated, mainly because of the influence of the body's 
rotation onto the gravitational field (related to the so--called
gravitomagnetic potential). Fortunately, for the exterior equations a powerful
solution technique (the `inverse scattering method' known from
soliton theory) exists in the case of axial symmetry and stationarity 
(\cite{mai} -- \cite{neu}). It
does not apply, unfortunately, to the interior equations. However, there is an
interesting limiting case: For infinitesimally thin disks the interior problem
`shrinks' to {\it boundary conditions}\/ for the exterior solution, and the 
solution technique mentioned can be utilized for solving the global problem.

A rigidly rotating disk of {\it dust}\/ is the universal limit of rigidly
rotating perfect fluid configurations as $p/\epsilon \to 0$ ($p$ denotes the
pressure and $\epsilon$ the mass--energy density), 
cf.~the disk limit of the Maclaurin
sequence. This disk is interesting for two reasons: On the one hand, it 
represents, in a sense, the simplest model of a 
self--gravitating rotating body (with no interaction except gravitation). On the 
other hand, it may serve as a crude model for astrophysical disks, for example
galaxies (with the stars considered as dust grains).  Of course, normal galaxies
are sufficiently well described by Newton's theory of gravitation. Nevertheless,
the relativistic model might be interesting, e.g.~in the context of quasars. 
An approximate solution was presented by Bardeen and Wagoner \cite{bw1}, 
\cite{bw2}.

The rigorous solution of the problem of the rigidly rotating disk of dust 
(\cite{nm1} -- \cite{nm3}) seems to provide the first example
of an exactly solvable rotating--body problem within general relativity, apart
from the Kerr solution describing a rotating black hole.

The lecture is organized as follows: In Section 2, a brief sketch of 
the method for solving boundary value 
problems of the axisymmetric, stationary vacuum Einstein equations is given.    
In section 3 the boundary value problem related to the dust disk and its
solution are discussed. The black hole limit of the solution is investigated
in some detail.

I would like to emphasize that most of the material of this lecture is based
on the joint work with Gernot Neugebauer and Andreas Kleinw\"achter 
(\cite{nm1} -- \cite{nkm}).
\section{Solution of boundary value problems to the axisymmetric, stationary
vacuum Einstein \newline equations}
\subsection{The linear system}
The axisymmetric, stationary vacuum Einstein equations (equivalent to
the so--called Ernst equation, see \cite{e}, \cite{kn}) are the 
{\it integrability condition}\/ of a related {\it linear system} (\cite{mai} --
\cite{neu}). Neugebauer's form \cite{neux} of the linear system reads
\begin{equation}
\Phi,_z=\left\{ \left( \begin{array}{cc} N & 0 \\ 0 & M \end{array} \right)
        +\lambda \left( \begin{array}{cc} 0 & N \\M & 0 \end{array} 
        \right) \right\}\Phi,
\label{Lin1}
\end{equation}
\begin{equation}
\Phi,_{\bar{z}}=\left\{ \left( \begin{array}{cc} \bar{M} & 0 \\ 0 & \bar{N} 
         \end{array} \right)
        +\frac{1}{\lambda} \left( \begin{array}{cc} 0 & \bar{M} \\ \bar{N} & 0 
        \end{array} \right) \right\}\Phi.
\label{Lin2}
\end{equation}
$\Phi(z,\bar{z},\lambda)$ is a $2\times 2$ -- matrix function depending on
\begin{equation}
z=\rho+i\zeta, \quad \bar{z}=\rho -i\zeta
\end{equation}
and
\begin{equation}
\lambda=\sqrt{\frac{K-i\bar{z}}{K+iz}}, 
\label{lam}
\end{equation}
where $K$ is an additional (complex) parameter, called the `spectral
parameter', which does not depend on the coordinates $\rho$ and $\zeta$.
($\rho$ and $\zeta$ are cylindrical coordinates, $\rho$ is the distance
to the symmetry -- [$\zeta$ --] axis.)
A bar denotes complex conjugation. The scalar functions $M$ and $N$ do not
depend on $\lambda$:
\begin{equation}
M=M(z,\bar{z}), \quad N=N(z,\bar{z}).
\end{equation}
The integrability condition 
\begin{equation}
\Phi,_{z\bar{z}}=\Phi,_{\bar{z}z}
\end{equation}
leads to a \underline{first order system}\footnote{
To obtain this system, one has to use the relations
\begin{displaymath}
\lambda,_z=\frac{\lambda}{4\rho}(\lambda^2-1), \quad \lambda,_{\bar{z}}=
\frac{1}{4\rho\lambda}(\lambda^2-1)
\end{displaymath}
following from (\ref{lam}). Comparing the coefficients of different powers of
$\lambda$ in the resulting equations (they must be valid for all $K$!) one  
obtains
\begin{displaymath}
N,_{\bar{z}}=N(\bar{M}-\bar{N})-\frac{1}{4\rho}(N+\bar{M}),\quad
M,_{\bar{z}}=M(\bar{N}-\bar{M})-\frac{1}{4\rho}(M+\bar{N})
\end{displaymath}
and the complex conjugate relations.
}
of nonlinear partial differential equations
for $M$, $N$, $\bar{M}$ and $\bar{N}$ which is equivalent to 
\begin{equation}
M=\frac{f,_z}{f+\bar{f}}, \quad N=\frac{\bar{f},_z}{f+\bar{f}}
\end{equation}
and the Ernst equation
\begin{equation}
(\Re f)(f,_{\rho \rho}+f,_{\zeta\zeta} + \frac{1}{\rho}f,_{\rho})
=f,_{\rho}^2+f,_{\zeta}^2
\label{ernst}
\end{equation}
for the complex function $f(\rho,\zeta)$, called the Ernst potential. 
(The relation to the metric can be found in section 3.1.) 

The existence of such a linear system with a spectral parameter
allows for the construction of exact solutions of the corresponding nonlinear
partial differential equation (here: the Ernst equation), e.g.~by means of
B\"acklund transformations \cite{ha}, \cite{neu}. These solutions contain an
arbitrary number of free parameters. More importantly, it is even possible to
construct solutions containing {\it free functions}. In this way, one can
solve, in principle, initial and/or boundary value problems. This method was
discovered by Gardner, Greene, Kruskal and Miura in 1967 \cite{ggkm} as a
method for solving the Cauchy problem of the Korteweg--de Vries (KdV) equation. 
The term `inverse scattering method' comes from the fact that one step of the
solution procedure consists in solving an inverse scattering problem for the
one--dimensional stationary Schr\"odinger equation which plays the role of
one part of the linear system related to the KdV equation.

The general idea behind is the discussion of the matrix function $\Phi$ as 
a function of the complex spectral parameter $K$ or, in our case, as a function
of $\lambda$. ($\rho$ and $\zeta$ play the role of parameters in this context.)
It is possible to obtain solutions containing free functions via
the solution of related Riemann--Hilbert problems in the complex 
$\lambda$--plane. This leads to linear integral equations, cf.~\cite{nmpz}.
\subsection{The Riemann--Hilbert technique}
A quite general
solution $\Phi$ of the linear system (\ref{Lin1}), 
(\ref{Lin2}) can be obtained by solving a matrix Riemann--Hilbert problem:
This problem consists in finding a $\Phi(\lambda)$ that is holomorphic for all
values of $\lambda$ in the complex plane 
except those which lie on some closed curves $\Gamma$ and
$\Gamma'$ defined by (\ref{lam}) and 
\begin{equation}
K \in \Gamma_K,
\end{equation}
with $\Gamma_K$ being a closed curve in the complex $K$--plane which is 
symmetric with respect to the real axis. (There exist two
curves in the $\lambda$--plane since $\lambda(K)$ is double--valued; $\lambda
\in \Gamma \Leftrightarrow -\lambda \in \Gamma'$.) 
On $\Gamma$ and $\Gamma'$ the following jump relations shall be satisfied:
\begin{equation}
\Phi_i=\Phi_e\,C(K) \quad \mbox{on $\Gamma$}, \quad
\Phi_i=\Phi_e\,C'(K) \quad \mbox{on $\Gamma'$},
\end{equation}
where $\Phi_i$ and $\Phi_e$ denote the values of $\Phi$ that appear by 
approaching the contour from inside and outside, respectively.  
It is assumed that the jump matrices $C(K)$ and $C'(K)$ do not depend
on $z$ and $\bar{z}$. As a consequence, the expressions $\Phi,_z\Phi^{-1}$
and $\Phi,_{\bar{z}}\Phi^{-1}$ {\it do not jump} on $\Gamma$ and $\Gamma'$.
Moreover, together with some additional 
assumptions in case of zeros of the determinant
${\rm det}\, \Phi(\lambda)$, one can show that $\Phi,_z\Phi^{-1}$ and
$\Phi,_{\bar{z}}\Phi^{-1}$ are holomorphic functions of $\lambda$ everywhere
except at the points $\lambda=\infty$ and $\lambda=0$, respectively.  
There, in agreement with (\ref{Lin1}) and (\ref{Lin2}), simple poles occur --
provided 
\begin{equation}
0 \not\in \Gamma, \quad \infty \not\in \Gamma.
\label{rev}
\end{equation}
Some constraints on the jump matrices $C$, 
$C'$ and a suitable normalization condition ensure 
that $\Phi,_z\Phi^{-1}$ and $\Phi,_{\bar{z}}\Phi^{-1}$ have exactly the 
structure as in (\ref{Lin1}), (\ref{Lin2}), and one can read off the $M$'s and
$N$'s or calculate the Ernst potential $f(\rho,\zeta)$. The solution of a matrix
Riemann--Hilbert problem can be found via a system of {\it linear} integral 
equations. In this way a solution of the Ernst equation is obtained which 
depends on free functions (some elements of the jump matrices which can be 
choosen arbitralily). Normally, this solution is regular 
for all values of $\rho$ and
$\zeta$ satisfying the condition (\ref{rev}), i.e., a curve $\Sigma$ in the 
$\rho$--$\zeta$--plane has to be excluded [cf.~(\ref{lam})]:
\begin{equation}
\Sigma: \quad \rho=|\Im K|, \quad \zeta=\Re K, \quad K \in \Gamma_K.
\label{sigma}
\end{equation}
This defines the surface of a body of revolution. One can try to solve
boundary value problems, e.g.~of the Dirichlet type,
with boundary data given on $\Sigma$.
The inverse scattering method would then consist of the following steps:
\begin{enumerate}
\item Determination of the jump matrices $C(K)$ and $C'(K)$ from the boundary
data. [The contour $\Gamma_K$ follows from the surface $\Sigma$ according
to (\ref{sigma})].
\item Solution of the Riemann--Hilbert problem.
\item Calculation of $f(\rho,\zeta)$ from $\Phi(z,\bar{z},\lambda)$.
\end{enumerate}
The first step is the most difficult one. It has to be solved by considering
the linear system (\ref{Lin1}), (\ref{Lin2}) along the boundary $\Sigma$.
(In the case of the application of the method to the solution of the Cauchy
problem of the KdV equation the first step is simpler and consists in solving
a `direct' scattering problem: One has to determine the `scattering data'
for a given potential.)

The second step consists in the solution of a system of linear integral
equations. (In the KdV case it corresponds to the inverse scattering problem:
One has to reconstruct a potential from scattering data. This leads to
the famous Gelfand--Levitan--Marchenko integral equation.)

The third step is trivial and provides us with the desired solution 
of the boundary value problem.

In the next section we discuss the problem of the rigidly rotating disk of dust,
which is the first example of a successful application of the 
procedure outlined. 
\section{The rigidly rotating disk of dust}
\subsection{The boundary value problem}
The disk of dust is characterized by the following energy--momentum tensor:
\begin{equation}
T^{ik}=\epsilon u^i u^k,\quad u^i=e^{-V}(\xi^i + \Omega \eta^i),
\label{ui}
\end{equation}
with  the mass--energy density $\epsilon$ and the four--velocity $u^i$. 
$\xi^i$ and $\eta^i$ are the Killing vectors corresponding to stationarity and
axisymmetry, respectively. The (positive) scalar $\exp(-V)$ follows 
from $u^iu_i=-1$, and $\Omega$ is the angular velocity as measured by an 
observer at infinity. Rigid rotation means
\begin{equation}
\Omega = constant.
\end{equation}
The line element can be written in the Weyl--Lewis--Papapetrou form
\begin{equation}
ds^2 = e^{-2U}[e^{2k}(d\rho^2+d\zeta^2) + \rho^2d\varphi^2] -
e^{2U}(dt + a\, d\varphi)^2,
\label{line}
\end{equation}
\begin{equation}
0\le\rho<\infty, \quad -\infty<\zeta<\infty,
\end{equation}
where $\exp(2U)$, $\exp(2k)$ and $a$ depend on $\rho$ and $\zeta$ only.
The Killing vectors, in these coordinates, are given by $\xi^i=\delta^i_t$,
$\eta^i=\delta^i_{\varphi}$. (Note, that we use units where the velocity of
light $c$ as well as Newton's gravitational constant $G$ are equal to 1.)
\begin{figure}[h]
\unitlength1cm
\begin{picture}(15,7)
\linethickness{1mm}
\put(0.5,2){\line(1,0){6}}
\thinlines
\put(6.5,2){\vector(1,0){6}}
\put(3.5,1){\vector(0,1){4}}
\put(3.4,5.4){$\zeta$}
\put(12.8,1.9){$\rho$}
\put(6.3,1.6){$\rho_o$}
\put(10.2,5){\underline{infinity:}}
\put(10.2,4.4){$f\rightarrow 1$}
\put(4,3.0){\underline{disk ($\Sigma$):}}
\put(4,2.4){$f'=e^{2V_o}$}
\end{picture}
\caption{The boundary value problem.}
\end{figure}
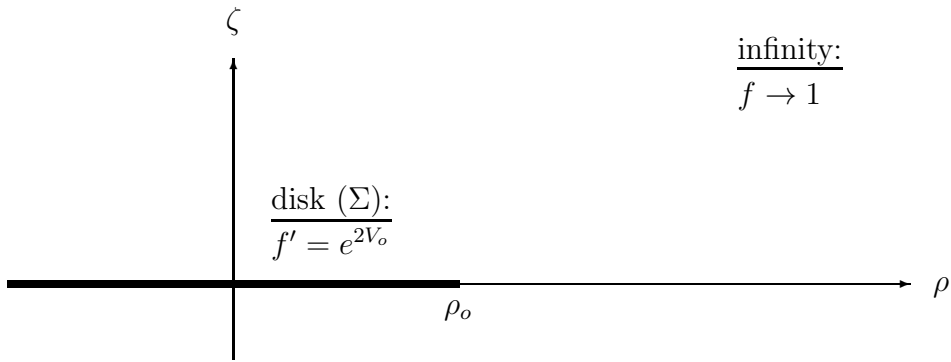
The disk is defined by $\zeta=0$, $\rho\le\rho_0$; $\rho_0$ being the 
(coordinate) radius of the disk. 
The mass density $\epsilon(\rho,\zeta)$ can be written formally as
\begin{equation}
\epsilon=\delta(\zeta)e^{U-k}\sigma_p(\rho),
\end{equation}
where $\delta(\zeta)$ is Dirac's delta function and $\sigma_p(\rho)$ is the 
(proper) surface mass--density.
The boundary conditions can easily be obtained by integrating the Einstein
equations across the disk, taking into account that the metric 
coefficients are
continuous, but their normal derivatives may jump. Together with the symmetry
of the problem with respect to the plane $\zeta=0$ one obtains
\begin{equation}
e^{2U'}=e^{2V_0}=constant, \quad a',_{\zeta}=0, \quad k',_{\zeta}=0 \quad
\mbox{as} \quad \zeta=0,\quad \rho\le\rho_0.
\label{bound}
\end{equation} 
The primed quantities refer to the corotating frame of reference defined by
\begin{equation}
\rho'=\rho,\quad \zeta'=\zeta,\quad \varphi'=\varphi - \Omega t,\quad t'=t.
\label{ko}
\end{equation}
The constant $V_0$ is related to the relative redshift $z_0$ measured by an
observer at infinity for photons coming from the center of the disk:
\begin{equation}
z_0=e^{-V_0}-1.
\end{equation}
[Note that the $V$ in (\ref{ui}) turns out to be constant and equal to $V_0$.]   
The surface mass--density can be calculated (after one has solved the 
global problem) by
\begin{equation}
\sigma_p=\left.\frac{1}{2\pi}e^{U-k}\,U',_{\zeta}\right|_{\zeta=0^+}.
\end{equation}
To formulate the global problem as a boundary value problem for the Ernst
equation, we need the connection of the metric functions with the
complex Ernst potential $f(\rho,\zeta)$:
\begin{equation}
f=e^{2U}+ib,
\label{deff}
\end{equation}
i.e., the real part of the Ernst potential is just equal to $\exp(2U)$.
The imaginary part $b$ is related to the gravitomagnetic potential $a$:
\begin{equation}
a,_{\rho}=\rho e^{-4U}b,_{\zeta},\quad a,_{\zeta}=-\rho e^{-4U}b,_{\rho}.
\label{ab}
\end{equation}
Hence $a(\rho,\zeta)$ can be calculated via a line integral from the Ernst 
potential and its derivatives. This integral is path--independent. [The 
integrability condition is satisfied as a consequence of (\ref{ernst}).]
Similarly, the metric function $k(\rho,\zeta)$ follows from 
\begin{equation}
k,_{\rho}=\rho[U,_{\rho}^2-U,_{\zeta}^2+
\frac{1}{4}e^{-4U}(b,_{\rho}^2-b,_{\zeta}^2)],\quad
k,_{\zeta}=2\rho(U,_{\rho}U,_{\zeta}+\frac{1}{4}e^{-4U}b,_{\rho}b,_{\zeta}).
\end{equation}
Note that, for regularity reasons, 
$a=0$ and $k=0$ on the symmetry axis ($\rho=0$), and we obtain, for example,
\begin{equation}
a=\int\limits_0^\rho \tilde{\rho} e^{-4U}b,_{\zeta} d\tilde{\rho},
\end{equation}
\begin{equation}
k=\int\limits_0^\rho \tilde{\rho}[U,_{\tilde{\rho}}^2-U,_{\zeta}^2+
\frac{1}{4}e^{-4U}(b,_{\tilde{\rho}}^2-b,_{\zeta}^2)] d\tilde{\rho}.
\end{equation}
[In the integrands, one has $U=U(\tilde{\rho},\zeta)$ and $b=b(\tilde{\rho},
\zeta)$.] 

Now, from (\ref{bound}), (\ref{deff}) and (\ref{ab}) we 
conclude that the boundary condition
for the `corotating' Ernst potential $f'$ is simply given by \cite{nm1}
\begin{equation}
f'=e^{2V_0}=constant \quad \mbox{as} \quad \zeta=0, \quad \rho\le\rho_0.
\label{b1}
\end{equation}
It should be noted that we have combined the result ${b'},_{\rho}=0$ 
and the freedom of adding an imaginary constant
to the Ernst potential, to set $b'=0$ in the disk.
The simple condition (\ref{b1}) for $f'$ in the 
corotating system 
corresponds to a quite complicated, \underline
{nonlocal boundary condition}\footnote{The coordinate transformation (\ref{ko})
preserves the form of the line element (\ref{line}). One obtains
\begin{displaymath}
e^{2U'}=e^{2U}[(1+\Omega a)^2 - \Omega^2\rho^2e^{-4U}],\quad
(1-\Omega a')e^{2U'}=(1+\Omega a)e^{2U},
\quad e^{2k'-2U'}=e^{2k-2U}.
\end{displaymath}
To get $b'$, one has to integrate the (primed) relations (\ref{ab}). 
That means, the boundary condition for $f$ is given by
complicated, nonlinear and nonlocal relations corresponding to $f'=e^{2V_0}$.
Note that these boundary conditions are equivalent to those applied by
Bardeen and Wagoner \cite{bw1}, \cite{bw2}. $V_0$ is identical with the
parameter $\nu_c$ of Bardeen and Wagoner.} 
for $f$ in the original system. On the other hand, the condition of asymptotic 
flatness is much simpler for $f$:
\begin{equation}
f\to 1 \quad \mbox{as} \quad \rho^2+\zeta^2\to\infty.
\label{as}
\end{equation} 
This condition ensures that $U\to 0$, $k\to 0$ and $a\to 0$ at infinity,
i.e., the line
element (\ref{line}) becomes Minkowskian (in cylindrical coordinates).
  
The global problem one has to solve, is to find a (or: {\it the}) 
solution of the Ernst
equation (\ref{ernst}) which satisfies (\ref{b1}) and (\ref{as}), and which 
is {\it regular everywhere outside the disk}, cf.~Figure 1.

The solution depends on two parameters only. One can choose, e.g., $V_0$ and
$\rho_0$, or $V_0$ and $\Omega$. A relation $\Omega=\Omega(V_0,\rho_0)$, see
next section, follows from the regularity condition at the rim of the disk. 
\subsection{The solution}
The solution was obtained by applying the method outlined in Section 2. The
curve $\Gamma_K$ in the complex $K$--plane is a part of the imaginary axis
(from $-i\rho_0$ to $+i\rho_0$) corresponding to the surface $\Sigma$, i.e.~the
disk, cf.~(\ref{sigma}) and Figure 1. 

The first step of the solution procedure
(the determination of the jump matrices) lead to the `small' integral equation
\cite{nm1} which could be solved in terms of elliptic functions \cite{nm2}.
The `big' integral equation \cite{nm1} corresponding to the second step
(the solution of the matrix Riemann--Hilbert problem) could be solved in terms 
of hyperelliptic functions and lead to the following result for the Ernst
potential \cite{nm3}:
\begin{equation}
f=\exp\left\{\int\limits_{K_1}^{K_a}\frac{K^2dK}{Z} +
\int\limits_{K_2}^{K_b}\frac{K^2dK}{Z} - v_2\right\},
\label{f}
\end{equation}
with
\begin{equation}
Z=\sqrt{(K+iz)(K-i\bar{z})(K^2-K_1^2)(K^2-K_2^2)},
\end{equation}
\begin{equation}
K_1=\rho_0\sqrt{\frac{i-\mu}{\mu}} \quad (\Re K_1<0), \quad K_2=-\bar{K}_1.
\label{K12}
\end{equation}
The real (positive) parameter $\mu$ is given by
\begin{equation}
\mu=2\Omega^2\rho_0^2e^{-2V_0}.
\label{mu}
\end{equation}
The upper integration limits $K_a$ and $K_b$ in (\ref{f})
have to be calculated from  
\begin{equation}
\int\limits_{K_1}^{K_a}\frac{dK}{Z} +
\int\limits_{K_2}^{K_b}\frac{dK}{Z} = v_0, \quad
\int\limits_{K_1}^{K_a}\frac{KdK}{Z} +
\int\limits_{K_2}^{K_b}\frac{KdK}{Z} = v_1,
\label{jacobi}
\end{equation}
where the functions $v_0$, $v_1$ and $v_2$ 
in (\ref{jacobi}) and (\ref{f}) are
given by
\begin{equation}
v_0=\int\limits_{-i\rho_0}^{+i\rho_0}\frac{H}{Z_1}dK,\quad
v_1=\int\limits_{-i\rho_0}^{+i\rho_0}\frac{H}{Z_1}KdK,\quad
v_2=\int\limits_{-i\rho_0}^{+i\rho_0}\frac{H}{Z_1}K^2dK,
\label{v}
\end{equation}
\begin{equation}
H=\frac{\mu\ln\left[\sqrt{1+\mu^2(1+K^2/\rho_0^2)^2} + \mu(1+K^2/\rho_0^2)
\right]}{\pi i \rho_0^2\sqrt{1+\mu^2(1+K^2/\rho_0^2)^2}} \quad (\Re H =0),
\end{equation}
\begin{equation}
Z_1=\sqrt{(K+iz)(K-i\bar{z})} \quad (\Re Z_1<0 \quad \mbox{for $\rho$, $\zeta$
outside the disk}).
\label{r}
\end{equation}
In (\ref{v}) one has to integrate along the imaginary axis. The 
integrations from $K_1$ to $K_a$ and $K_2$ to $K_b$ in (\ref{f}) and 
(\ref{jacobi}) have to 
be performed along the same paths in the two--sheeted Riemann surface associated
with $Z(K)$. The problem of finding $K_a$ and $K_b$ from 
(\ref{jacobi}) is a special case of Jacobi's inversion problem. 
It generalizes the
definition of elliptic functions and can be solved in terms of hyperelliptic
theta functions (\cite{goe}, \cite{ro}, see also \cite{st} -- \cite{bob}). 
Using a formula for Abelian integrals of the third kind derived by Riemann
(see \cite{st}) it is also possible to express the Ernst potential $f$ directly
in terms
of theta functions \cite{nkm}. On the symmetry axis ($\rho=0$) and in the plane
of the disk ($\zeta=0$) all integrals in (\ref{f}) and (\ref{jacobi}) 
are reduced to elliptic ones \cite{nm2}.

The solution (\ref{f}) satisfies the boundary conditions (\ref{b1}) and
(\ref{as}), has a positive surface mass-- (particle number--) density (vanishing
at the rim of the disk),
and it is regular everywhere outside the disk -- provided
\begin{equation}
0<\mu<\mu_0=4.62966184\dots
\end{equation}
(for $\mu>\mu_0$ one or more singular rings appear in the plane $\zeta=0$,
outside the disk). The interesting behaviour for $\mu\to\mu_0$ will be discussed
in the next section.

Note that the solution in the form (\ref{f}) -- (\ref{r}) depends on the 
parameters $\rho_0$ and $\mu$ only. Since $\exp(2U')=\exp(2U)$ on the symmetry
axis ($\rho=0$), one can calculate the parameter $V_0$ [cf.~(\ref{bound})]
from $\Re f(\rho=0,\zeta=0^+)$. The result is \cite{nm2}:
\begin{equation}
V_0=-\frac{1}{2}\sinh ^{-1}\left\{\mu+\frac{1+\mu^2}
{\wp[I(\mu);\frac{4}{3}\mu^2-4,\frac{8}{3}\mu(1+\mu^2/9)]-\frac{2}{3}\mu}
\right\},
\label{V0}
\end{equation}
\begin{equation}
I(\mu)=\frac{1}{\pi}\int\limits_0^{\mu}\frac{\ln(x+\sqrt{1+x^2})dx}
{\sqrt{(1+x^2)(\mu-x)}}
\end{equation}
($\wp$ is the \underline{Weierstra\ss\ function}\footnote{The Weierstra\ss\ 
function $\wp(x;g_2,g_3)$ is defined by
\begin{displaymath}
\int\limits_{\wp(x;g_2,g_3)}^{\infty}
\frac{dt}{\sqrt{4t^3-g_2t-g_3}}=x.
\end{displaymath}}),
i.e., $V_0$ depends on $\mu$ alone. The range $0<\mu<\mu_0$ corresponds to
$0>V_0>-\infty$. In this range, the relation (\ref{V0}) can be
inverted uniquely to give $\mu(V_0)$. [$\mu_0$ is the first zero of the denominator
in (\ref{V0}).]  
Then, from the definition (\ref{mu}) one  
obtains the relation $\Omega(V_0,\rho_0)$. (Without loss of generality, we
assumed $\Omega>0$; the solution for negative $\Omega$ is simply given
by $\bar{f}$.) Alternatively, one can use $V_0$ and $\Omega$ as the primary
parameters, with $\rho_0=\rho_0(V_0, \Omega)$.
\subsection{Discussion: \newline 
>From the Newtonian limit to the black hole limit} 
For $0<\mu<\mu_0$, the solution (\ref{f}) can be expanded in 
terms of $\mu^{1/2}$:
\begin{equation}
f=1+\sum\limits_{n=1}^{\infty}f_n\,\mu^{(n+1)/2} = 1 + f_1\,\mu +
{\cal O}(\mu^{3/2}).
\end{equation}
The $f_n$ ($n=1,2,\dots,\infty$) are elementary functions of $\rho$ and $\zeta$.
(This series corresponds to the 
\underline{Bardeen--Wagoner expansion}\footnote{The expansion parameter $\gamma$
used by Bardeen and Wagoner \cite{bw1}, \cite{bw2} is related to $\mu$ by
$\gamma = 1-e^{V_0(\mu)}=\mu/2+{\cal O}(\mu^2)$.}.) 
The Newtonian limit ($\mu\ll 1$) is represented
by $f_1$:
\begin{equation}
f_1=-\frac{1}{\pi}\left\{\frac{4}{3}\cot^{-1}\xi +
[\xi-(\xi^2+\frac{1}{3})\cot^{-1}\xi](1-3\eta^2)\right\},
\end{equation}
with elliptic coordinates $\xi$ and $\eta$:
\begin{equation}
\rho=\rho_0\sqrt{1+\xi^2}\sqrt{1-\eta^2},\quad \zeta=\rho_0\xi\eta \quad
(0\le\xi<\infty,\,\, -1\le\eta\le 1)
\end{equation}
($\xi=0$ is the disk). The Newtonian potential $U_N$ is obtained from
\begin{equation}
g_{tt}=-\Re f =-(1+f_1\,\mu) + {\cal O}(\mu^{3/2}) = -(1+\frac{2U_N}{c^2})
+{\cal O}(\frac{1}{c^3}),
\label{N}
\end{equation}
i.e.,
\begin{equation}
U_N=\Omega^2\rho_0^2\,f_1.
\end{equation}
This is exactly the Maclaurin solution. [Note that we have reintroduced the
velocity of light $c$ into Eq.~(\ref{N}). From (\ref{mu}) and (\ref{V0}) 
one obtains $\mu=2\Omega^2\rho_0^2/c^2 + {\cal O}(\mu^2)$.]

With increasing $\mu$, characteristic deviations from the Newtonian solution
occur. This concerns, e.g., the radial distribution of the surface mass--density
\cite{bw2}, \cite{nm2}. 
The gravitomagnetic potential leads to dragging effects and,
for $\mu>1.68849\dots$, even to the formation of an ergoregion \cite{bw2},
\cite{mk}. Some
illustrations can also be found in \cite{nkm}.

However, the most striking difference to the Newtonian case is 
the following \cite{bw2}, \cite{nm1}:
 
For given angular momentum the mass of the disk is bounded by
\begin{equation}
\frac{M^2}{J}<1,
\end{equation}
where $M$ and $J$ denote the total (gravitational) mass and the 
($\zeta$--component of the) total angular momentum, 
respectively ($c=G=1$ again). The equality
$M^2/J=1$ is reached in the limit $\mu\to\mu_0$, the {\bf black hole limit}.

For $\mu\to\mu_0$, one has $V_0\to-\infty$, cf.~(\ref{V0}). 
As a consequence of (\ref{mu}), for nonvanishing $\Omega$, this results in
\begin{equation}
\rho_0\to 0.
\end{equation}
For $\rho^2+\zeta^2\ne 0$, we obtain from (\ref{f}) in the limit $\mu\to\mu_0$
\begin{equation}
f=\frac{2\Omega r-1-i\cos \vartheta}{2\Omega r +1-i\cos \vartheta} \qquad
(r > 0), 
\label{ek}
\end{equation} 
\begin{equation}
\rho=r\sin \vartheta, \quad \zeta=r\cos \vartheta \quad
(0\le\vartheta\le\pi).
\end{equation}
This is exactly the ($r>0$ part of the) 
\underline{extreme Kerr solution}\footnote{
To derive (\ref{ek}) from (\ref{f}), let us first rewrite (\ref{f}) and 
(\ref{jacobi}) in the equivalent form
\begin{displaymath}
f=\exp\left\{\int\limits_{K_b}^{K_a}\frac{K^2dK}{Z} - \tilde{v}_2\right\},
\quad \int\limits_{K_b}^{K_a}\frac{dK}{Z}=\tilde{v}_0, \quad
\int\limits_{K_b}^{K_a}\frac{KdK}{Z}=\tilde{v}_1,
\end{displaymath}
with
\begin{displaymath}
\tilde{v}_n=v_n - \int\limits_{K_1}^{K_2}\frac{K^ndK}{Z} \quad (n=0,1,2).
\end{displaymath}
($K_b$ is now on the other sheet of the Riemann surface.) In the limit 
$\mu\to\mu_0$ one obtains for $r>0$, using (\ref{V0}),
\begin{displaymath}
\tilde{v}_0=\frac{2\Omega}{r} - \frac{\pi i\cos\vartheta}{2r^2}, 
\quad \tilde{v}_1=-\frac{\pi i}{2r}, \quad \tilde{v}_2=0
\end{displaymath}
(modulo periods). 
In the above integrals from $K_b$ to $K_a$, because of $K_1\to 0$, $K_2\to 0$
[cf.~(\ref{K12})],
$Z$ can be replaced by $Z=K^2\sqrt{(K+iz)(K-i\bar{z})}$. Hence, all integrals 
become elementary and the result (\ref{ek}) can easily (and uniquely) 
be obtained.
} 
with
\begin{equation}
M=\frac{1}{2\Omega}\, ,\quad J=\frac{1}{4\Omega^2}\, ,
\end{equation}
i.e.,
\begin{equation}
\frac{M^2}{J}=1.
\end{equation}
(Remember that we assumed $\Omega>0$.)
In the coordinates used, the horizon of the extreme Kerr black hole is 
just given
by (the excluded) $r=0$. $\Omega$ plays the role of the `angular velocity of
the horizon'. In the corotating system (\ref{ko}), one obtains
\begin{equation}
f'=-\Omega^2r^2\left[\frac{2(1+i\cos\vartheta)^2}{2\Omega r +1-i\cos \vartheta}
+\sin^2\vartheta\right]. 
\label{ek'}
\end{equation}
It can be seen, that the boundary condition (\ref{b1}) with $V_0\to -\infty$
is indeed satisfied on the horizon ($r=0$). 

A completely different limit of the space--time, for $\mu\to\mu_0$, 
is obtained for finite values of $r/\rho_0$ (corresponding
just to the previously excluded $r=0$). Therefore, we consider a coordinate
transformation \cite{bw2}
\begin{equation}
\tilde{r}=r\,e^{-V_0}, \quad \tilde{\varphi}=\varphi-\Omega t, \quad
\tilde{\vartheta}=\vartheta, \quad \tilde{t}=t\,e^{V_0}.
\end{equation} 
(Note that finite $r/\rho_0$ correspond to finite $\tilde{r}$ in the limit.)
For $\mu<\mu_0$, this is nothing but the transformation to the corotating
system (\ref{ko}) combined with a rescaling of $r$ and $t$. The 
transformed Ernst potential $\tilde{f}$ is related to $f'$ according to
$\tilde{f}=f'\exp(-2V_0)$, i.e.,
\begin{equation}
\frac{\tilde{f}}{\tilde{r}^2}=\frac{f'}{r^2} \quad \mbox{as} \quad \mu<\mu_0.
\label{ff}
\end{equation}
However, for $\mu\to\mu_0$, the solutions $f'$ (finite $r$) 
and $\tilde{f}$ (finite $\tilde{r}$) separate from each other. 
(A similar phenomenon has been observed by Breitenlohner {\it et al.} for
some limit solutions of the static, spherically symmetric 
Einstein--Yang--Mills--Higgs equations \cite{bfm}.) 
For finite $r$,
the extreme Kerr solution (\ref{ek'}) arises, while finite $\tilde{r}$ lead to
a solution which still describes a disk. This solution (which can
be expressed in terms of theta functions) is regular
everywhere outside the disk, but it is {\it not asymptotically flat}, i.e., it
can be considered as a cosmological solution. The space--time structure of
both solutions ($f'$ and $\tilde{f}$) coincides at $r=0$ (the horizon)
and $\tilde{r}\to\infty$ (spatial infinity). The relation (\ref{ff}) survives
in the form
\begin{equation}
\lim\limits_{\tilde{r}\to\infty}\,\frac{\tilde{f}}{\tilde{r}^2}=
\lim\limits_{r\to 0}\,\frac{f'}{r^2} \quad \mbox{as} \quad \mu\to\mu_0.
\end{equation}
Accordingly, for $\mu\to\mu_0$ and $\tilde{r}\to\infty$,
\begin{equation}
\tilde{f}\to\tilde{f}_{as} = -\Omega^2\tilde{r}^2\left[
\frac{2(1+i\cos\tilde{\vartheta})^2}{1-i\cos \tilde{\vartheta}}
+\sin^2\tilde{\vartheta}\right].
\end{equation}
Note that $\tilde{f}_{as}$ belongs to the family of solutions to
the Ernst equation of the type $f=r^kY_k(\cos \vartheta)$ 
presented by Ernst \cite{e2}. The corresponding 
{\it asymptotic} line element is given by
the following exact solution of the vacuum Einstein equations:
\begin{equation}
ds^2=e^{-2\tilde{U}}[e^{2\tilde{k}}(d\tilde{r}^2+\tilde{r}^2
d\tilde{\vartheta}^2)+\tilde{r}^2\sin^2\tilde{\vartheta}\,d\tilde{\varphi}^2]
-e^{2\tilde{U}}(d\tilde{t}+\tilde{a}\,d\tilde{\varphi})^2\quad ,
\end{equation}
\begin{equation}
e^{2\tilde{U}}=\Omega^2\tilde{r}^2\cdot\frac{\cos^4\tilde{\vartheta}+
6\cos^2\tilde{\vartheta}-3}{\cos^2\tilde{\vartheta}+1}\quad ,
\end{equation}
\begin{equation}
\tilde{a}=\frac{2}{\Omega^2\tilde{r}}\cdot\frac{\cos^2\tilde{\vartheta}-1}
{\cos^4\tilde{\vartheta}+6\cos^2\tilde{\vartheta}-3}\quad ,
\end{equation}
\begin{equation}
e^{2\tilde{k}}=\frac{1}{4}(\cos^4\tilde{\vartheta}+6\cos^2\tilde{\vartheta}-3)
\, .
\end{equation}
These analytical results prove the conjectures formulated by Bardeen and Wagoner 
\cite{bw2} on the basis of their numerical results. 

Let me conclude with a
quotation from Bardeen and Wagoner (\cite{bw2}, page 411): 
{\it `The picture we have developed of the extreme relativistic limit of
a rotating disk is that it becomes buried in the horizon of the $J=M^2$ Kerr
metric, surrounded by its own infinite, non asymptotically flat universe. As
approached from the $r>0$ Kerr region, the disk represents a ``singularity''
in the horizon, since the whole range $0\le\tilde{r}<\infty$, over which
there exist considerable changes in the local geometry, corresponds to an
infinitesimal range of affine parameter for a typical photon which reaches 
the horizon from the outside.'}


\begin{thebibliography}{99} 
\bibitem{li} L.~Lichtenstein, {\it Gleichgewichtsfiguren rotierender
Fl\"ussigkeiten} (Springer 1933)
\bibitem{cha} S.~Chandrasekhar, {\it Ellipsoidal figures of equilibrium}
(Yale University Press 1969)
\bibitem{bw1} J.M.~Bardeen and R.V.~Wagoner, 
Astrophys.~J.~{\bf 158} (1969) L65
\bibitem{bw2} J.M.~Bardeen and R.V.~Wagoner, 
Astrophys.~J.~{\bf 167} (1971) 359
\bibitem{nm1} G.~Neugebauer and R.~Meinel, Astrophys.~J.~{\bf 414} (1993) L97
\bibitem{nm2} G.~Neugebauer and R.~Meinel, 
Phys.~Rev.~Lett.~{\bf 73} (1994) 2166
\bibitem{nm3} G.~Neugebauer and R.~Meinel, 
Phys.~Rev.~Lett.~{\bf 75} (1995) 3046
\bibitem{mn1} R.~Meinel and G.~Neugebauer, Class.~Quantum Grav.~{\bf 12} 
       (1995) 2045
\bibitem{mn2} R.~Meinel and G.~Neugebauer, Phys.~Lett.~{\bf 210 A} (1996) 160

\bibitem{kmn} A.~Kleinw\"achter, R.~Meinel and G.~Neugebauer, 
       Phys.~Lett.~{\bf 200 A} (1995) 82
\bibitem{mk} R.~Meinel and A.~Kleinw\"achter, Einstein Studies (Birkh\"auser)
       {\bf 6} (1995) 339
\bibitem{nkm} G.~Neugebauer, A.~Kleinw\"achter and R.~Meinel, Helv.~Phys.~Acta
{\bf 69} (1996) 472  
\bibitem{e} F.J.~Ernst, Phys.~Rev.~{\bf 167} (1968) 1175
\bibitem{kn} D.~Kramer and G.~Neugebauer, Commun.~Math.~Phys.~{\bf 7} (1968) 173
\bibitem{mai} D.~Maison, Phys.~Rev.~Lett.~{\bf 41} (1978) 521
\bibitem{bz} V.A.~Belinski and V.E.~Zakharov, Zh.~Eksper.~Teoret.~Fiz.~Pis'ma
{\bf 75} (1978) 195
\bibitem{ha} B.K.~Harrison, Phys.~Rev.~Lett.~{\bf 41} (1978) 119
\bibitem{neu} G.~Neugebauer, J.~Phys.~{\bf A 12} (1979) L67, 
{\bf A 13} (1980) L19
\bibitem{neux} G.~Neugebauer, J.~Phys.~{\bf 13 A} (1980) 1737
\bibitem{ggkm} C.S.~Gardner, J.M.~Greene, M.D.~Kruskal and R.M.~Miura,
Phys.~Rev.~Lett. {\bf 19} (1967) 1095
\bibitem{nmpz} S.~Novikov, S.V.~Manakov, L.P.~Pitaevskii and V.E.~Zakharov,
{\it Theory of Solitons} (Consultants Bureau 1984)
\bibitem{goe} A.~G\"opel, Crelle's J.~f\"ur Math.~{\bf 35} (1847) 277
\bibitem{ro} G.~Rosenhain, Crelle's J.~f\"ur Math.~{\bf 40} (1850) 319
\bibitem{st} H.~Stahl, {\it Theorie der Abel'schen Funktionen} (Teubner 1896)
\bibitem{kra} A.~Krazer, {\it Lehrbuch der Thetafunktionen} (Teubner 1903)
\bibitem{bob} E.D.~Belokolos, A.I.~Bobenko, V.Z.~Enol'skii, A.R.~Its and
V.B.~Matveev, {\it Algebro-Geometric Approach to Nonlinear Integrable
Equations} (Springer 1994) 
\bibitem{bfm} P.~Breitenlohner, P.~Forg\'acs and D.~Maison, 
Nucl.~Phys.~{\bf 442 B} (1995) 126
\bibitem{e2} F.J.~Ernst, J.~Math.~Phys.~{\bf 18} (1977) 233    
\end{thebibliography}
\end{document}